\title{A Note on Parity Violation and the Immirzi Parameter}
\author{
        Andrew Randono \\
               Center for Relativity, Department of Physics\\
        University of Texas at Austin \\
        Austin, TX 78712\\
                  email: arandono@physics.utexas.edu
}

\documentclass[12pt]{article}

\begin{document}
\maketitle
\bibliographystyle{utphys}

\begin{abstract}
There has been considerable recent interest in the Immirzi parameter as a measure of parity violating effects in the
classical theory of gravitation with fermion coupling. Most recently it was shown that the Immirzi parameter
together with the non-minimal coupling constant of Dirac spinors provides the measure for parity violating
spin-spin interaction terms in the effective field theory. For complex values of the Immirzi parameter, the resulting
effective field theory yields complex values for the torsion, and a non-unitary effective field theory that blows up
for the special cases $\gamma=\pm i$ where the gravitational kinetic term is the Ashtekar action. We show that by
restricting ourselves to real values for the torsion, there
is a natural set of choice for the non-minimal coupling
constant that yields real and unitary effective field theory that does not blow up for the
special cases $\gamma=\pm i$. We then show that these particular values for the non-minimal coupling coefficients
most naturally follow from a non-minimal pseudo-kinetic term in the fermion Lagrangian.
\end{abstract}

\section{Introduction}
It has recently been shown that the gravitational Immirzi term together with the non-minimal fermion coupling yields
parity violating effects in the effective field theory \cite{Rovelli:Torsion},\cite{Freidel:Torsion}. The calculation
yields several peculiar features
for imaginary values of the Immirzi parameter. First, the contorsion tensor is complex yielding complex values for the
torsion. This suggests that the calculation only holds when either the gauge group is complexified, the tangent bundle
is complexified, or both. Second, the effective field theory blows up for the special cases $\gamma=\pm i$ where the
connection is (anti)self-dual. Tracing back the root if this infinite term one finds that the source is the inverse of
the
matrix $\frac{1}{2}\left(1+i\frac{1}{\gamma}\gamma_{5}\right)$ which is  non-invertible for the special values $\gamma=\pm
i$ where it is the right(left) chiral projection operator. 

We show that for the case of any imaginary Immirzi parameter, by restricting to real values for the torsion there is
only one natural choice (barring Majorana-type fermions) for the non-minimal coupling constant $\alpha$, which
is not independent of the Immirzi parameter, namely $\alpha=\frac{1}{\gamma}$ for left-handed fermions and
$\alpha=-\frac{1}{\gamma}$ for right-handed fermions. The resulting field theory is finite and unitary. It does not
blow up for the special cases $\gamma=\pm i$, and has the same form as that of the Einstein-Cartan theory with
$\alpha=0$ and $\gamma=0$.

The need for different species-dependent non-minimal coupling constants may appear unsatisfying from a theoretical
perspective, especially when one attempts to include non-chiral fermions. Thus, we show that the effective field theory
most naturally follows from the Dirac Lagrangian together with an equivalent non-minimal pseudo-kinetic term with one
universal coupling constant $\beta=\frac{1}{\gamma}$. It is, of course, possible to start with this Lagrangian from
the beginning, but for logical clarity we first follow closely the derivations in \cite{Freidel:Torsion}, and derive the
equivalent action.

\section{Real Torsion}
Throughout this
paper we will assume the veirbein $e^{I}$ and the $Spin(3,1)$ connection coefficients $\omega^{IJ}$ are
real. The veirbein, being an invertible map from a spin-1 vector representation space $V$ to the cotangent bundle
$T^{*}M$ must be real so long as the cotangent bundle is real. Requiring the coefficients $\omega^{IJ}$ to be real
is tantamount to choosing a real gauge group. If we now split the connection into a torsion free part and the
contorsion tensor
\begin{equation}
{\omega_{\mu I}}^{J}={\omega(e)_{\mu I}}^{J}+{C_{\mu I}}^{J}
\end{equation}
we see that the contorsion tensor must also be real. Thus, when solving for the
contorsion tensor from the minimization of the Holst action together with non-minimally coupled fermions we will search
only for real solutions. Surprisingly, this will force the non-minimal coupling constant $\alpha$ to be imaginary for
imaginary values of $\gamma$, as opposed to the previously derived real values.

The initial procedure follows exactly analogous to \cite{Freidel:Torsion}. We begin with the Holst action,
\begin{equation}
S_{H}=\frac{1}{16\pi G}\int d^{4}x e e^{\mu}_{I}e^{\nu}_{J}{P^{IJ}}_{KL}{R_{\mu\nu}}^{KL}
\end{equation} 
where
\begin{equation}
{P^{IJ}}_{KL}=\frac{1}{2}\left(\delta^{IJ}_{KL}-\frac{1}{\gamma}{\epsilon^{IJ}}_{KL}\right)
\end{equation}
together with the non-minimally coupled Dirac kinetic term
\begin{equation}
S_{F}=\int d^{4}x\frac{i e}{2}e^{\mu}_{I}\left((1-i\alpha)\overline{\psi}\gamma^{I}\nabla_{\mu}\psi-(1+i\alpha)
\overline{\nabla_{\mu}\psi}\gamma^{I}\psi \right).
\end{equation}
As pointed out in \cite{Freidel:Torsion}, the variation of the action gives different equations of motion for $\psi$ and
$\overline{\psi}$ unless the following condition holds:
\begin{equation}
(\alpha-\alpha^{*})e^{\mu}_{J}{C_{\mu}}^{IJ}. \label{condition1}
\end{equation}
This can be solved by taking $\alpha$ to be real, but it is automatically satisfied if ${C_{\mu}}^{I\mu}=0$. The
equations of motion we will find for \textit{real} values of ${C_{\mu}}^{IJ}$ automatically
satisfy this condition, so alpha can (and must) be complex. The equation we must solve is the same as before,
and is obtained by varying the action with respect to the connection coefficients ${\omega_{\mu I}}^{J}$:
\begin{equation}
\frac{1}{2\pi G}\left({C_{KL}}^{Q}+{C_{\nu K}}^{\nu}\delta^{Q}_{L}\right){P^{KL}}_{IJ}=
{\epsilon^{QL}}_{IJ} A_{L}+2\alpha \delta^{Q}_{[I}V_{J]}. \label{master}
\end{equation} 
Since we are looking for real solutions to the contorsion tensor, it will be useful to solve this
equation by dividing it up into real and imaginary parts as opposed to inverting ${P^{KL}}_{IJ}$. Furthermore, the
matrix $P$ is non-invertible for the special values $\gamma=\mp i$ where $\frac{1}{2}P$ is the left(right) chiral
projection operator. So, let us now break up $\alpha$ into its real and imaginary parts,
$\alpha=\alpha_{r}+i\alpha_{i}$ where $Re(\alpha)=\alpha_{r}$ and $Im(\alpha)=\alpha_{i}$, 
and begin by solving the imaginary part of (\ref{master}) given by: 
\begin{equation}
\frac{1}{2\pi G}\left({C_{KL}}^{Q}+{C_{\nu K}}^{\nu}\delta^{Q}_{L}\right){\epsilon^{KL}_{IJ}}=
4 s(\gamma)|\gamma|\alpha_{i}\delta^{Q}_{[I}V_{J]}.
\end{equation}
where we have written $\gamma=i\ s(\gamma)|\gamma|$ (that is, $s(\gamma)$ is the sign of $\gamma$ assuming it is pure
imaginary).
By inverting the alternating symbol and making the appropriate contractions one can easily show
\begin{equation}
{C_{\nu I}}^{\nu}=0 \label{trace}
\end{equation}
from which follows
\begin{equation}
{C_{[MN]}}^{Q}=-2\pi G (s(\gamma)|\gamma|\alpha_{i})\left({\epsilon_{MN}}^{QJ}V_{J} \right). \label{torsion1}
\end{equation}
To relate this directly to the torsion, we recall ${C_{[\mu\nu]}}^{\rho}=-\frac{1}{2}{T_{\mu\nu}}^{\rho}$.
We now need to determine if this solution is compatible with the real part of (\ref{master}) given by:
\begin{equation}
\frac{1}{2\pi G}\left({C_{[IJ]}}^{Q}+{C_{\nu [I}}^{\nu}\delta^{Q}_{J]}\right)=
{\epsilon_{IJ}}^{QK}A_{K}+2\alpha_{r}\left(\delta^{Q}_{[I}V_{J]}\right). \label{realpart}
\end{equation} 
Contracting the $Q$ and the $I$ indices we find 
\begin{equation}
{C_{QJ}}^{Q}=-\alpha_{r}3 V_{J}.
\end{equation}
For this to be compatible with (\ref{trace}) without severely restricting the
spinor fields, we must have $\alpha_{r}=0$. Thus, we see that condition (\ref{condition1}) is satisfied not because
$\alpha$ is real (it is imaginary), but because ${C_{\nu I}}^{\nu}=0$.
Inserting this back into (\ref{realpart}) gives:
\begin{equation}
{C_{[MN]}}^{Q}=2\pi G \left({\epsilon_{MN}}^{QJ}A_{J}\right). \label{torsion2} 
\end{equation}
For the two equations for the torsion, (\ref{torsion1}) and (\ref{torsion2}), to be compatible we must have
\begin{eqnarray}
A_{J} &=& \alpha\gamma V_{J}. \label{condition2}
\end{eqnarray}
This rather peculiar equation can be solved by by recalling for a left(right) handed spinor field, $A_{J}=\pm
V_{J}$. Thus, we solve (\ref{condition2}) by setting $\alpha=\pm \frac{1}{\gamma}$ and demanding $\psi=\psi_{L}$ if
$\alpha=\frac{1}{\gamma}$ and $\psi=\psi_{R}$ if $\alpha=-\frac{1}{\gamma}$.\footnote{It may be possible to solve
(\ref{condition2}) trivially by introducing Majorana-type fermions. We will not consider Majorana-type fermions in this
paper but we leave it as an open possibility}

\section{The Unitary and Finite Effective Field Theory}
We now repeat the procedure in \cite{Freidel:Torsion} to construct the effective field theory. The calculations here are
considerably simpler because the contorsion tensor is much simpler. Recalling the form of the contorsion tensor in
terms of the torsion:
\begin{eqnarray}
{C_{[MN]}}^{Q} &=& -\frac{1}{2}{T_{MN}}^{Q} \\
{C_{MN}}^{Q} &=& \frac{1}{2}\eta^{QR}\left[T_{MRN}+T_{NRM}-T_{MNR}\right]
\end{eqnarray}
we solve for the contorsion tensor
\begin{equation}
{C_{MN}}^{Q}=(2\pi G) {\epsilon_{MN}}^{QK}A_{K}. \label{contorsion}
\end{equation}

The first current-current interaction term comes from the torsion part of the Einstein-Cartan action
\begin{equation}
S^{(1)}=\frac{1}{16\pi G}\int d^{4}x e e^{\mu}_{I}e^{\nu}_{J}{P^{IJ}}_{KL}[C_{\mu},C_{\nu}]^{KL}.
\end{equation}
Using the identity
\begin{equation}
e^{\mu}_{I}e^{\nu}_{J}{P^{IJ}}_{KL}[C_{\mu},C_{\nu}]^{KL}=6(U^{2}-\frac{2}{\gamma}UW-W^{2})
\end{equation}
if $C_{IJK}=\epsilon_{IJKL}U^{L}+2\eta_{I[J}V_{K]}$.
This yields a total contribution:
\begin{equation}
S^{(1)}=\frac{3}{2}\pi G\int d^{4}x e A^{2}.
\end{equation}
The second contribution
comes from the torsion-current interaction terms given by\footnote{Here we have corrected a
missing factor of 2 from the original changing the $\frac{1}{8}$ to $\frac{1}{4}$}
\begin{equation}
S^{(2)}=\int d^{4}x e \frac{1}{4}C^{IJK}\left(\epsilon_{IJKL}A^{L}+2\alpha \eta_{I[J}V_{K]}\right).
\end{equation}
This yields a total contribution
\begin{equation}
S^{(2)}=-3\pi G \int d^{4}x e A^{2}.
\end{equation}
The total current-current interaction in the effective field theory is then given by
\begin{eqnarray}
S_{int}=S^{(1)}+S^{(2)} &=& -\frac{3}{2}\pi G\int d^{4}x e A^{2} \\
&=& -\frac{3}{2}\pi G \int d^{4}x e (\alpha \gamma)^{2}V^{2}. \label{interaction}
\end{eqnarray}

For consistency, we now check that these solutions agree with those of \cite{Freidel:Torsion}. There it was assumed that
$\alpha$ was pure real, however, this fact was not used in most of the derivations so we can safely let $\alpha$ be
imaginary. We first consider their solution for the contorsion:
\begin{equation}
C_{IJK}=4 \pi G \frac{\gamma^{2}}{1+\gamma^{2}}\left(\frac{1}{2}\epsilon_{IJKL}
(A+\frac{\alpha}{\gamma}V)^{L}-\frac{1}{\gamma}\eta_{I[J}(A-\alpha\gamma V)_{K}\right). \label{contorsion2}
\end{equation}
We now demand that the contorsion $C_{IJK}$ is real. One can easily show that if $\gamma$ is real and $\alpha$ is
imaginary or vice versa (but both are non-zero and finite), then either $A^{I}$ or $V^{I}$ are zero. Now consider the
case when both $\gamma$ and $\alpha$ are imaginary. Then, provided that the contorsion is real, (\ref{contorsion2})
yields
\begin{equation}
\eta_{I[J}(\frac{1}{\gamma}A-\alpha V)_{K]}=0
\end{equation}
whose solution is our condition (\ref{condition2}):
\begin{equation}
A^{I}=\alpha\gamma V^{I}.
\end{equation}
Inserting this back into (\ref{contorsion2}) gives our solution (\ref{contorsion}) for the contorsion. The interaction
terms in \cite{Freidel:Torsion} are given by:
\begin{equation}
S_{int}=-\frac{3}{2}\pi G \frac{\gamma^{2}}{1+\gamma^{2}}\int d^{4}x e 
\left(A^{2}-\frac{2\alpha}{\gamma}AV-\alpha^{2}V^{2}\right). \label{FreidelInteraction}
 \end{equation}
Inserting $A=\alpha\gamma V$ into this yields precisely our solution (\ref{interaction}) for the interaction. Crucial
in the derivation is the cancellation of the factor $\gamma^{2}/(1+\gamma^{2})$. Without this factor, the solution is
well defined for all imaginary values of the Immirzi parameter including the special cases $\gamma=\pm i$.
Furthermore, the interaction term is real, so it will not violate unitarity. We would like to point out that our
effective field theory has the same form as the effective field theory of ordinary Einstein-Cartan theory as can be
easily obtained by setting $\alpha=0$ and taking the limit as $\gamma$ goes to infinity in
(\ref{FreidelInteraction}).
The only difference is that we have the additional chiral constraint (\ref{condition2}). However, it is well known that
the non-perturbative quantum theories are very different due to the presence of the Immirzi parameter.

\section{The equivalent non-minimal pseudo-kinetic action}
From a theoretical perspective, it may appear unsatisfying to require different values for $\alpha$ depending on the
chirality of the fermion. This objection is especially poignant when one considers the coupling of non-chiral
fermions. There, one must break the non-chiral spinor up into its left and right components and add non-minimal
coupling terms with coefficients $\alpha=\frac{1}{\gamma}$ and $\alpha=-\frac{1}{\gamma}$ respectively. Since one
usually thinks of a non-chiral fermion, say an electron or any massive fermion, as a single entity, adding different
coupling coefficients to its left and right-handed parts may seem a bit ad-hoc.

One can resolve this issue by considering the equivalent non-minimal pseudo-kinetic term
\begin{equation}
S'_{NM}=-i\beta\int d^{4}x \frac{ie}{2}e^{\mu}_{I}\left(\overline{\psi}\gamma_{5}\gamma^{I}\nabla_{\mu}\psi
+\overline{\nabla_{\mu}\psi}\gamma_{5}\gamma^{I}\psi\right).
\end{equation}
Setting $\beta$ universally equal to $\frac{1}{\gamma}$ yields the same classical theory described above. To see this,
simply decompose $\psi$ into its right and left parts and use the identities $\gamma_{5}\psi_{L}=-\psi_{L}$,
$\gamma_{5}\psi_{R}=\psi_{R}$:
\begin{eqnarray}
S'_{NM} &=& \int d^{4}x  \frac{e}{2} e^{\mu}_{I}
 \ \frac{1}{\gamma}\left(\overline{\psi_{L}}
\gamma^{I}\nabla_{\mu}\psi_{L} +\overline{\nabla_{\mu}\psi_{L}}\gamma^{I}\psi_{L}\right)\\
& &  -\frac{1}{\gamma}\left(\overline{\psi_{R}}\gamma^{I}\nabla_{\mu}\psi_{R} 
+\overline{\nabla_{\mu}\psi_{R}}\gamma^{I}\psi_{R}\right).
\end{eqnarray}
\section*{Acknowledgments}
Many thanks to Laurent Freidel for discussions of his work with D. Minic and T. Takeuchi and commentary on the issue of
imaginary $\gamma$.
\bibliography{ParityV1}

\providecommand{\href}[2]{#2}\begingroup\raggedright\begin{thebibliography}{1}

\bibitem{Rovelli:Torsion}
A.~Perez and C.~Rovelli, ``Physical effects of the {I}mmirzi parameter,''
  \href{http://www.arXiv.org/abs/arXiv:gr-qc/0505081 v2}{{\tt
  arXiv:gr-qc/0505081 v2}}.

\bibitem{Freidel:Torsion}
L.~Freidel, D.~Minic, and T.~Takeuchi, ``Quantum gravity, torsion, parity
  violation and all that,'' \href{http://www.arXiv.org/abs/arXiv:hep-th/0507253
  v2}{{\tt arXiv:hep-th/0507253 v2}}.

\end{thebibliography}\endgroup

\end{document}